\documentclass[%
 reprint,
 superscriptaddress,
 showpacs,preprintnumbers,
 amsmath,amssymb,
 aps,
 prl,
 longbibliography,
lengthcheck,%
]{revtex4-1}

\usepackage{graphicx}% Include figure files
\usepackage{dcolumn}% Align table columns on decimal point
\usepackage{bm}% bold math
\usepackage{color}
\usepackage{ulem}
\usepackage{hyperref}% add hypertext capabilities
\usepackage{xr}
\begin{document}

\preprint{APS/123-QED}

\title{
Nonlinear optical response of truly chiral phonons:\\
Light-induced phonon angular momentum, Peltier effect, and orbital current
%Generation of phonon angular momentum, spin current, and nonlinear Peltier effect in truly chiral materials with electromagnetic waves
}

\author{Hiroaki Ishizuka}
\affiliation{
Department of Physics, Institute of Science Tokyo, Meguro, Tokyo, 152-8551, Japan
}

\author{Masahiro Sato}
\affiliation{
Department of Physics, Chiba University, Chiba 263-8522, Japan
}

\date{\today}

\begin{abstract}
The nonlinear optical responses of chiral phonons to terahertz and infrared light are studied using the nonlinear response theory.
We show that the photo-induced angular momentum increases with the square of the chiral-phonon relaxation time $\tau$, giving a significantly larger angular momentum compared to ordinary phonons.
We also find that the photo-induced Peltier effect by chiral phonons occurs through a mechanism distinct from those proposed recently; the induced energy current scales $\propto\tau^2$, giving a larger energy current in the clean limit.
We prove a linear relation between the generated angular momentum and the energy current. 
Lastly, we show that the orbital current, an analog of the spin current, occurs through a nonlinear response.
These findings demonstrate the unique properties and functionalities of chiral phonons.
\end{abstract}
% Generate angular momentum > light-induced Bernet effect. 

\pacs{
}% PACS, the Physics and Astronomy

\maketitle

%%%%   Introduction   %%%%

\begin{figure}
  \includegraphics[width=\linewidth]{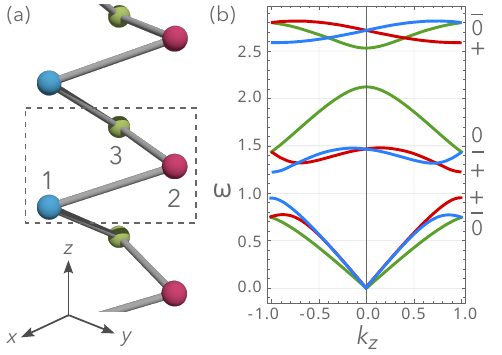}
  \caption{
  (a) Schematic of Te-like three-ion model. The dashed box indicates the unit cell. (b) The phonon dispersion of Te-like model. The red, green, and blue curves are the bands with crystal angular momentum $L^z=1$, $0$, and $-1$, respectively. 
  }\label{fig:model}
\end{figure}

{\it Introduction} ---
Chiral phonon is a phonon mode whose angular momentum and the wavenumber couples, $\bm L\cdot\bm k\ne0$, where $\bm L$ is the crystal angular momentum and $\bm k$ is the momentum of phonons, respectively~\cite{Ishito2023a}.
In chiral crystals, the degeneracy of chiral phonons is lifted, and the modes possess a non-zero group velocity at the symmetric points in the Brillouin zone [Fig.~\ref{fig:model}(b)]~\cite{Kishine2020a}.
The coupling of the group velocity and the chirality allows small-momentum phonons to carry energy and angular momentum efficiently, providing unique properties and making them attractive for applications.

In studying phonons and exploring their functionalities, the optical response has played a central role.
For example, Raman scattering techniques were used to study truly chiral phonons~\cite{Ishito2023a}, %~\cite{Ishito2023a,Zhang2024a}, \tr{( replace or remove [2])} 
and optical generation of phonon angular momentum and related phenomena were studied in several materials~\cite{Zhu2018a,Juraschek2020a,Juraschek2022a,Geilhufe2023a,Li2023a,Basini2024a,Davies2024a,Choi2024a,Lagarde2024a,Tong2024a}.
As demonstrated by these works, unlike phononic heat conduction and thermal Hall effect~\cite{Strohm2005a,Sheng2006a,Inyushkin2007a,Sheng2006a,Kagan2008a,Zhang2010a,Mori2014a,Saito2019a,Zhang2019a}, which are often dominated by acoustic gapless phonons, optical techniques are especially useful for studying and utilizing optical gapped phonons.
On the other hand, recent studies on the photocurrent of electrons revealed intriguing features, such as non-local current transport~\cite{Ishizuka2017a,Nakamura2017a}, robustness against disorder~\cite{Morimoto2018a,Harada2020a,Ishizuka2021a}, and the generation of spin~\cite{Young2013a} and orbital~\cite{Davydova2022a} photocurrents.
Similar phenomena in magnetic excitations~\cite{Proskurin2018a,Ishizuka2019a,Ishizuka2019b,Ishizuka2022a,Fujiwara2025a} and phonons~\cite{Ishizuka2024a} were theoretically predicted.
%pointed out.
Chiral phonons have properties distinct from the models used in these studies, possibly realizing unique nonlinear responses.
%However, the understanding of such nonlinear responses, especially those unique to chiral phonons, is highly limited.
%
%The unique features of chiral phonons may also affect nonlinear responses to electromagnetic waves.
%The nonlinear response is ...
%Recently, it has been pointed out that such nontrivial transport phenomena may also occur in magnons and phonons.
%These phenomena offer a new route to realizing novel functionalities in the infrared and THz regime.

To understand the optical properties of chiral phonons, we study a series of second-order nonlinear responses to terahertz and infrared light: Angular momentum generation, photo-induced nonlinear Peltier effect, and orbital current.
For the angular momentum generation and photo-induced Peltier effect, we find that a mechanism unique to the chiral phonons induces a larger response than that in non-chiral phonons. 
The angular momentum and phonon photocurrent induced by this mechanism are polarization-dependent, distinct from those induced by linearly polarized light.
In addition, we argue that an orbital photocurrent of phonons occurs in chiral materials, which might be functional for spintronics applications.
%We argue that a chirality-dependent phonon photocurrent occurs by truly chiral phonons.
%The phonon photocurrent occurs by the circularly polarized light, with the direction of heat current ... 

{\it Chiral phonon} --- An example of the model with truly-chiral phonon modes and its dispersion are shown in Fig.~\ref{fig:model}~\cite{Tsunetsugu2023a}.
The model is an one-dimensional atom chain consisting of a lattice with three ions in a unit cell forming a spiral structure.
The Hamiltonian reads,
\begin{align}
&\hat H_{CP}=\sum_{ia\mu}\frac{\hat p^2_{ia\mu}}{2m_a}\nonumber\\
&+\frac12\sum_{(ia\mu,jb\nu)}(\hat u_{ia\mu}-\hat u_{jb\mu})D_{ia\mu,jb\nu}(\hat u_{ia\nu}-\hat u_{jb\nu}),\label{eq:H}
\end{align}
where $D_{ia\mu,jb\nu}$ is the dynamical matrix, $\hat u_{ia\mu}$ is the displacement along the $\mu$ axis of $a\,(=1,2,3)$ sublattice atom in $i$th unit cell, and $p_{ia\mu}$ is the conjugate momentum of $\hat u_{ia\mu}$ satisfying $[\hat u_{ia\mu},\hat p_{jb\nu}]={\rm i}\delta_{ij}\delta_{ab}\delta_{\mu\nu}$; here we set the Dirac constant $\hbar=1$ and the sum is over all pairs of $ia\mu$.
The dynamical matrix is $D_{i-1\,3\mu,i1\nu}=(D_{31})_{\mu\nu}$ where 
\begin{align}
D_{31}=\left(\begin{array}{ccc}K_{11}&0&0\\0&K_{22}&-\Delta K\\0&-\Delta K&K_{33}\\\end{array}\right),
\end{align}
for the nearest-neighbor bonds between $a=1$ and $3$ sublattices, and the nearest-neighbor matrices for the other two bonds are given by $D_{31}$ rotated by 120$^\circ$. 
Here, $K_{ii}$ ($i=1,2,3$) and $\Delta K$ are the stiffness constants~\cite{Tsunetsugu2023a}.

The phonon bands in Fig.~\ref{fig:model}(b) are characterized by the crystal angular momentum of ions~\cite{Tsunetsugu2023a}; the phonon bands with non-zero crystal angular momentum are called chiral phonons.
A unique feature of truly chiral phonons is the linear crossing of phonon bands at the symmetric point [$\Gamma$ in the case of Fig.~\ref{fig:model}(b)].
The crossing of two phonon bands results from different angular momentum, which prohibits the anti-crossing.
In the below, we focus on doubly degenerate cases in which only two bands cross at $\bm k=\bm0$.

{\it Nonlinear response theory} ---
To study the second-order dc response of quantum phonons to electromagnetic waves, we start by considering non-interacting phonon models with $n_{ph}$ phonon bands.
The Hamiltonian is given by
\begin{align}
\hat H=\sum_{\substack{\bm k,n}}\omega_{n\bm k}(\hat b_{n\bm k}^\dagger\hat b_{n\bm k}+\frac12),
\end{align}
and the light-matter coupling,
\begin{align}
\hat H'=&\sum_{\mu,n,\bm k}[\beta^\mu_{n\bm k}\hat b_{n\bm k}+(\beta^\mu_{n\bm k})^\ast\hat b_{n\bm k}^\dagger] E_\mu(t),\label{eq:H'}
\end{align}
with $\beta^\mu_{n\bm k}$ being the coupling constant and $E^\mu(t)$ is the external field.
Here, $\hat b_{n\bm k}$ ($\hat b_{n\bm k}^\dagger$) is the annihilation (creation) operator of a phonon with band index $n$ and momentum $\bm k$, and $\omega_{n\bm k}$ ($\omega_{n\bm k}\le \omega_{m\bm k}$ if $n<m$) is its phonon frequency.
%$\tau=\tau(T)$ is the phenomenological phonon lifetime at temperature $T$, and $o^\lambda_{nm}(\bm k)$ are the matrix elements of the physical quantity,
The observables we study in this work are assumed to be 
\begin{align}
\hat O_\lambda=&\sum_{n,m,\bm k}\hat b_{n\bm k}^\dagger o^\lambda_{nm}(\bm k)\hat b_{m\bm k}+\hat b_{n\bm k}^\dagger o^\lambda_{n\bar m}(\bm k)\hat b_{m-\bm k}^\dagger\nonumber\\
&+\hat b_{n-\bm k} o^\lambda_{\bar nm}(\bm k)\hat b_{m\bm k}+\hat b_{n-\bm k} o^\lambda_{\bar n\bar m}(\bm k)\hat b_{m-\bm k}^\dagger.\label{eq:O}
\end{align}
The general formula of the nonlinear response tensor $\chi_{\lambda;\mu\nu}(\Omega;\omega,\Omega-\omega)$ for phonons are given recently~\cite{Ishizuka2024a}, which is also summarized in the supplemental material~\cite{suppl}.
Here, $\Omega$ and $\omega$ $(-\omega)$ are, respectively, the frequencies of the observable and the $\mu$ $(\nu)$ component of the applied wave.

For the case of the direct coupling between the ion charge and a uniform electric field, which we focus on, $\beta^\mu_{n\bm k}$ reads
\begin{align}
\beta_{n\bm 0}^{\mu} =&\sum_{a}q_a|n\bm 0\rangle_{a\mu} \sqrt{\frac{ N}{2m_a\omega_{n\bm 0}}},\label{eq:coupling}
\end{align}
and $\beta_{n\bm k}^{\mu} =0$ for $\bm k\ne\bm 0$.
Here, $N$ is the number of unit cells, $|n\bm k\rangle$ is the $n$th eigenstate of dynamical matrix with the wavenumber $\bm k$, and $q_a$ is the charge of ions on the $a$th sublattice.

\begin{figure}
  \includegraphics[width=\linewidth]{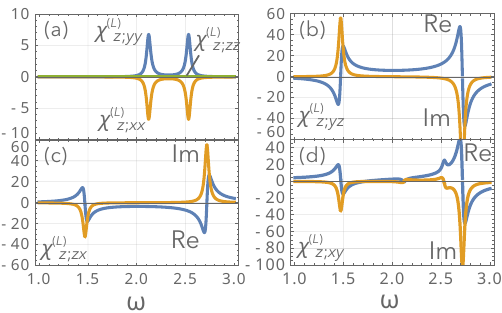}
  \caption{
  (a) Real part of the nonlinear susceptibility $\chi^{(L)}_{z;xx}(0;\omega,-\omega)$, $\chi^{(L)}_{z;yy}(0;\omega,-\omega)$, and $\chi^{(L)}_{z;zz}(0;\omega,-\omega)$, and the frequency dependence of (b) $\chi^{(L)}_{z;yz}(0;\omega,-\omega)$, (c) $\chi^{(L)}_{z;zx}(0;\omega,-\omega)$, and (d) $\chi^{(L)}_{z;xy}(0;\omega,-\omega)$ for the Te-like model. The results are for $\delta=0.23$, $K_1=3/2$, $K_2=1/2$, $K_3=3$, $\theta=
 0.3\pi$, $a=1$, $c=1$, $m=1$, $q_1=0$, $q_2=1$, $q_3=-1$, and $\tau=20$.
  }\label{fig:chi}
\end{figure}

{\it Phonon angular momentum} --- 
Using the equilibrium position of ion $r_{ia\mu}^0=r_{ia\mu}-u_{ia\mu}$ where $r_{ia\mu}$ is the position of the atom, the angular momentum of a lattice reads $\bm L^{\rm atm}=\sum_{ia}\bm r_{ia}\times\bm p_{ia}=\bm L^{\rm lat}+\bm L^{\rm ph}$ where $\bm L^{\rm lat}=\sum_{ia}\bm r_{ia}^0\times\bm p_{ia}$ is the lattice angular momentum and $\bm L^{\rm ph}=\sum_{ia} \bm u_{ia}\times\bm p_{ia}$ is the phonon angular momentum. 
The $z$ component of the phonon (crystal) angular momentum is given by Eq.~\eqref{eq:O},
%\begin{align} 
%\hat L_z^{\rm ph}&=\sum_{n,m,\bm k}\hat b_{n\bm k}^\dagger M_{nm}(\bm k)\hat b_{m\bm k}+\sum_{n,m,\bm k}\hat b_{n\bm k}^\dagger M_{n\bar m}(\bm k)\hat b_{m-\bm k}^\dagger\nonumber\\
%&+\sum_{n,m,\bm k}\hat b_{n-\bm k} M_{\bar nm}(\bm k)\hat b_{m\bm k}+\sum_{n,m,\bm k}\hat b_{n-\bm k} M_{\bar n\bar m}(\bm k)\hat b_{m-\bm k}^\dagger,
%\end{align}
where $o^\lambda_{nm}(\bm k)$ is replaced by $M_{nm}(\bm k)=\frac14\frac{\omega_{m\bm k}+\omega_{n\bm k}}{\sqrt{\omega_{n\bm k}\omega_{m\bm k}}}\langle n\bm k|M^z|m\bm k\rangle=-M_{\bar n\bar m}(\bm k)$ and $M_{\bar nm}(\bm k)=-\frac14\frac{\omega_{n\bm k}-\omega_{m\bm k}}{\sqrt{\omega_{n\bm k}\omega_{m\bm k}}}\langle n\bm k|M^z|m\bm k\rangle=-M_{n\bar m}(\bm k)$~\cite{Bozovic1984a,Zhang2014a,Chen2019a,suppl}.
Here, $M^z$ is a matrix whose elements are $(M^\lambda)_{a\mu,b\nu}=-i\delta_{ab}\epsilon_{\lambda\mu\nu}$, %$I_{\rm uc}\otimes\left(\sigma^y\oplus 0_1\right)$, where $0_1$ is the $1\times1$ zero matrix, $\sigma^y=\left(\begin{array}{cc}0&-i\\i&0\\\end{array}\right)$, and $I_{\rm uc}$ is the $n_{\rm uc}\times n_{\rm uc}$ unit matrix with $n_{\rm uc}$ being the number of ions in the unit cell; $I_{\rm uc}$ acts on the ion subspace and $\sigma^y\oplus 0_1$ on the direction.
where $\delta_{ab}$ is the Kronecker's delta and $\epsilon_{\mu\nu\lambda}$ is the Levi-Civita tensor.

In view of the optical responses, the phonon angular momentum generated by light irradiation can be described as a second-order dc response.
The $\omega$ dependence of the nonlinear response tensor $\chi^{(L)}_{\lambda;\mu\nu}(0;\omega,-\omega)$ for the angular momentum is shown in Fig.~\ref{fig:chi}.
In Figs.~\ref{fig:chi}(b), the two peaks of ${\rm Im}[\chi^{(2)}_{z;xy}(0;\omega,-\omega)]$ at $\omega\simeq1.47$ and $2.72$, which correspond to the frequencies of $\bm k=\bm0$ chiral phonons in Fig.~\ref{fig:model}(b), indicate that the angular momentum of phonons is generated by exciting chiral phonons with circularly polarized light. %[Fig.~\ref{fig:chi}(d)].
This is consistent with the intuition that circularly polarized light induces the circular motion of atoms.
Figures~\ref{fig:chi}(b) and \ref{fig:chi}(c) also show that the angular momentum is generated by the circularly polarized electric field in the $yz$ and $zx$ planes. 
%as shown in Fig.~\ref{fig:chi}(b) and \ref{fig:chi}(c) \tr{(b) and (d)?}.
The results indicate that angular momentum can be induced with circularly polarized light regardless of the incident direction in the Te-like model.

%Next, we discuss the relaxation time dependence of the phonon angular momentum.
%As shown in the inset of Fig.~\ref{fig:chi}(a), 
We note that the height of the peaks at the chiral phonon frequencies increases ${\rm Im}[\chi^{(2)}_{z;\mu\nu}(0;\omega,-\omega)]\propto\tau^2$ in the long $\tau$ limit, unlike the usual resonance peaks that increase as $\tau^1$. 
The peaks for polarization-dependent angular momentum appears from the $\omega_{n\bm0}=\omega_{m\bm0}$ terms in the general formula (Eq.~\eqref{suppl:eq:chi_full} in~\cite{suppl}), which reads
\begin{align}
\chi^{(L)}_{\lambda;\mu\nu}\sim
&\frac{2\tau}{\pi V}\sum_{\alpha,\beta}^{N_n}\frac{\beta_{n\alpha\bm 0}^\mu\Im[M_{n\alpha,n\beta}(\bm 0)](\beta_{n\beta\bm 0}^\nu)^\ast}{\omega-\omega_{n\bm 0}-\frac{\rm i}{2\tau}}, %\nonumber\\
%&-\frac{2\tau}{\pi V}\sum_{\alpha,\beta}^{N_n}\frac{(\beta_{n\alpha\bm 0}^\mu)^\ast\Im[M_{n\alpha,n\beta}(\bm 0)]\beta_{n\beta\bm 0}^\nu}{\omega+\omega_{n\bm 0}-\frac{\rm i}{2\tau}},
\end{align}
around $\omega=\omega_{n\bm0}$.
Here, we relabeled the bands by a pair of indexes $n\alpha$, where $n$ is for different energy levels $\omega_n(\bm0)$ and $\alpha$ is the index for $N_n$ degenerate bands with frequency $\omega_n(\bm0)$.
The equation shows that the height of the peak increases $\propto\tau^2$.
In addition, for the quadratic phonon models in Eq.~\eqref{eq:H}, one can take the eigenstates as $|n\alpha\bm k\rangle=|n\alpha,-\bm k\rangle^\ast$, in which case the coefficients of the angular momentum operator satisfies $M_{n\alpha,m\beta}(\bm k)=M_{\overline{m\beta},\overline{n\alpha}}(-\bm k)=-M_{m\beta,n\alpha}(-\bm k)$;
in addition, $\beta_{n\alpha\bm 0}^\nu$ in Eq.~\eqref{eq:coupling} becomes a real number.
Combined with hermiticity, $M_{n\alpha,m\beta}(\bm k)=M_{m\beta,n\alpha}^\ast(\bm k)$, the above equality gives $M_{n\alpha,n\alpha}(\bm0)=0$ and $M_{n\alpha,n\beta}(\bm0)$ purely imaginary for $\alpha\ne\beta$, indicating that the $\tau^2$ peaks appear only if the degenerate bands exist and that the $\tau^2$ response is dependent on polarization.

The $\tau^2$ dependence suggests that these contributions show behaviors distinct from the well-known resonance phenomena.
For instance, they (the $\tau^2$ terms) may dominate the nonlinear response to electromagnetic waves in a material with a long lifetime, realizing a large phonon angular momentum. %\tr{(can we remove the following?)} provided that the series expansion remains valid to the clean limit.
Such a situation is likely for phonons, whose lifetime is typically longer than that of electronic and magnetic excitations. 
%Experimentally, the optically induced phonon angular momentum is known to be very large~\cite{Juraschek2017a,Zhu2018a,Zhou2025a}. 
As chiral phonons, not limited to truly chiral phonons, are degenerate at symmetric points, the $\tau^2$ dependence may explain the large angular momentum~\cite{Basini2024a}. 

In the last, as shown in Fig.~\ref{fig:chi}(a), we note that a finite angular momentum can also be induced by linearly-polarized light.
In the figure, two peaks appear in $\chi^{(L)}_{z;xx}$ and $\chi^{(L)}_{z;yy}$ at $\omega=2.12$ and $2.53$, the frequencies corresponding to non-chiral phonons ($L_z^{\rm ph}=0$ phonons).
%It shows that a finite angular momentum occurs by the irradiation of linearly polarized light.
%it is somewhat different from those obtained by transferring photon angular momentum to the phonons; in such a case, the incident direction of the electromagnetic waves is parallel to the direction of angular momentum (the $z$ axis in the current case).
However, the peaks of the response function increase as $\propto \tau^1$ for these non-chiral phonons. 
Generating angular momentum using linearly polarized light may appeal to the generation of phonon angular momentum with low-frequency phonons, at which frequency generating a high-intensity circularly polarized light is challenging.

%
% \tau^2 > explain the large angular momentum induced by the light
% rotation of electron?
%

{\it Peltier effect of phonons} --- 
Another characteristic of truly chiral phonons is the non-zero group velocity at $\bm k=\bm0$~\cite{Kishine2020a}, facilitating efficient phonon transport.
An interesting phenomenon in this context is the optical Peltier effect of phonons~\cite{Ishizuka2024a}, which is a temperature gradient induced by a photocurrent of phonons.
The coefficient for the Peltier effect, $\Pi_{\lambda;\mu\nu}(0;\omega,-\omega)$, is calculated by replacing $o^\lambda$ with the phonon energy current operator~\cite{Hardy1963a},
$J^\lambda_{nm}(\bm k)=\frac{\omega_{n\bm k}+\omega_{m\bm k}}2v_{nm}^\lambda(\bm k)$, $J^\lambda_{n\bar m}(\bm k)=\frac{\omega_{n\bm k}-\omega_{m\bm k}}2v_{nm}^\lambda(\bm k)$, $J^\lambda_{\bar n\bar m}(\bm k)=-J^\lambda_{nm}(\bm k)$, $J^\lambda_{\bar nm}(\bm k)=-J^\lambda_{n\bar m}(\bm k)$, where
%\begin{align}
  $v^\lambda_{nm}(\bm k)=\frac{\langle n\bm k|\partial_{k_\lambda}\tilde A(\bm k)|m\bm k\rangle}{4\sqrt{\omega_{n\bm k}\omega_{m\bm k}}}$
%\end{align}
is the phonon velocity operator with $\tilde A_{a\mu,b\nu}(\bm k)=\sum_i\frac{A_{ia\mu,0b\nu}}{\sqrt{m_am_b}}e^{-{\rm i}\bm k\cdot(\bm r_{ia}-\bm r_{0b})}$.
Here, we expressed the Hamiltonian in the form $H_{CP}=\frac12\sum_{ia\mu,jb\nu}A_{ia\mu,jb\nu}u_{ia\mu}u_{jb\nu}$ for the sake of conciseness~\cite{Hardy1963a,Ishizuka2024a}; $A_{ia\mu,jb\nu}$ are $A_{ia\mu,ia\mu}=\sum_{jb\nu}D_{ia\mu,jb\nu}$ and  $A_{ia\mu,jb\nu}=-D_{ia\mu,jb\nu}$ for the Hamiltonian in Eq.~\eqref{eq:H}.

\begin{figure}
  \includegraphics[width=\linewidth]{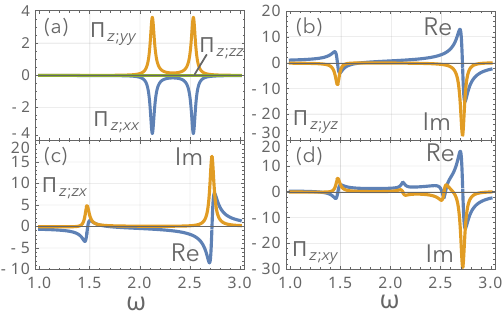}
  \caption{
  Real part of the Peltier coefficient for (a) $\Pi_{z;xx}(0;\omega,-\omega)$, $\Pi_{z;yy}(0;\omega,-\omega)$, and $\Pi_{z;zz}(0;\omega,-\omega)$, and the frequency dependence of (b) $\Pi_{z;yz}(0;\omega,-\omega)$, (c) $\Pi_{z;zx}(0;\omega,-\omega)$, and (d) $\Pi_{z;xy}(0;\omega,-\omega)$ for the Te-like model. The results are for $\delta=0.23$, $K_1=3/2$, $K_2=1/2$, $K_3=3$, $\theta=
 0.3\pi$, $a=1$, $c=1$, $m=1$, $q_1=0$, $q_2=1$, $q_3=-1$, and $\tau=20$.
  }\label{fig:pi}
\end{figure}

Figure~\ref{fig:pi} shows the $\omega$ dependence of the Peltier coefficient for the Te-like model.
The resonance peaks for $\Pi_{z;xx}$ and $\Pi_{z;yy}$ appear at the frequencies for the non-chiral phonons as shown in Fig.~\ref{fig:pi}(a), similar to those in a previous work~\cite{Ishizuka2024a}.
In contrast, the resonance peaks appear in the imaginary part for the chiral phonons, $\omega\simeq1.47$ and $2.72$;
they indicate that a polarization-dependent Peltier effect occurs with circularly polarized light [Figs.~\ref{fig:pi}(b)-\ref{fig:pi}(d)].

To further investigate the nature of the polarization-dependent Peltier effect, we look into the general formula (Eq.~\eqref{suppl:eq:chi_full}) assuming a uniform external field, i.e., $\beta_{n\bm k}^{\mu} =0$ for $\bm k\ne\bm 0$.
Similar to the case of angular momentum, $\Pi_{\lambda;\mu\nu}$ near the resonance frequency, $\omega=\omega_{n\bm0}$, reads
\begin{align}
\Pi_{\lambda;\mu\nu}\sim
&\frac{2\tau}{\pi V}\sum_{\sigma,\sigma'}^{N_n}\frac{\beta_{n\sigma\bm 0}^\mu\Im[v^\lambda_{n\sigma,n\sigma'}(\bm 0)](\beta_{n\sigma'\bm 0}^\nu)^\ast}{\omega-\omega_{n\bm 0}-\frac{\rm i}{2\tau}}. %\nonumber\\
%&-\frac{2\tau}{\pi V}\sum_{\sigma,\sigma'}^{N_n}\frac{(\beta_{n\sigma\bm 0}^\mu)^\ast\Im[v^\lambda_{n\sigma,n\sigma'}(\bm 0)]\beta_{n\sigma'\bm 0}^\nu}{\omega+\omega_{n\bm 0}-\frac{\rm i}{2\tau}}.
\label{eq:pi21}
\end{align}
%Here, we relabeled the bands by a pair of indexes $n\sigma$, where $n$ is for different energy levels $\omega_n(\bm0)$ and $\sigma$ is the index for $N_n$ degenerate bands with frequency $\omega_n(\bm0)$.
In addition, $v^\lambda_{n\sigma,m\sigma'}(\bm k)=v^\lambda_{\overline{m\sigma'}\overline{n\sigma}}(-\bm k)=-v^\lambda_{m\sigma',n\sigma}(-\bm k)$ and $v^\lambda_{n\sigma\overline{m\sigma'}}(\bm k)=v^\lambda_{m\sigma'\overline{n\sigma}}(-\bm k)$ holds for an eigenstate basis that satisfies $|n\sigma\bm k\rangle=|n\sigma,-\bm k\rangle^\ast$; the property also holds for the photo-induced angular momentum. 
%, in which case the coefficients of phonon energy current operator satisfies $v^\lambda_{n\sigma,m\sigma'}(\bm k)=v^\lambda_{\overline{m\sigma'}\overline{n\sigma}}(-\bm k)=-v^\lambda_{mn}(-\bm k)$, $v^\lambda_{n\bar m}(\bm k)=v^\lambda_{m\bar n}(-\bm k)$, $v^\lambda_{\bar nm}(\bm k)=v^\lambda_{\bar mn}(-\bm k)$.
%The first equality gives $v_{n\sigma,n\sigma}^\lambda(\bm0)=0$, indicating that the contribution from non-degenerate bands vanishes~\cite{Ishizuka2024a}.
%Hence, only the degenerate phonon bands at the $\Gamma$ point contribute to the polarization-dependent Peltier effect.
Hence, as in the case of the angular momentum, only the degenerate phonon bands contribute to the polarization-dependent Peltier effect.

Equation~\eqref{eq:pi21} also shows that the $\tau$ dependence of the peak height scales as $\Pi\propto\tau^2$ in the large $\tau$ limit, which dependence is distinct from the relation $\Pi\propto\tau$ for the linearly polarized light~\cite{Ishizuka2024a}.
Therefore, for a clean material at low temperatures, the polarization-dependent Peltier effect should be larger. 
%induce a larger Peltier effect.
The different $\tau$ dependence arises from the property of $v^\lambda_{n\sigma,m\sigma'}(\bm k)$ discussed in the previous paragraph, which gives $v^\lambda_{n,n}(\bm 0)=0$.
Hence, the leading order term in Eq.~\eqref{eq:pi21} vanishes and the subleading contribution with $\tau^1$ dependence becomes the dominant contribution.

The difference in $\tau$ dependence resembles that of the shift and injection currents in photovoltaics~\cite{vBaltz1981a,Sipe2000a}, and the spin current by two-magnon processes~\cite{Ishizuka2019a}.
However, we note that the $\tau$ dependence is $\tau^0$ and $\tau^1$ for shift and injection currents, respectively.
This power-law difference arises from the absence of the integral over $\bm k$ in the photocurrent.
In the case of photovoltaic effects, an integral over the momentum exists as the current occurs through the excitation of electrons with an arbitrary momentum in the Brillouin zone.
%If the integral over $\bm k$ exists in Eq.~\eqref{eq:pi21},
In such a case, the integral transforms the Lorenzian in the denominator of Eq.~\eqref{eq:pi21} into a constant, reducing the order of $\tau$ by one.
From Eq.~\eqref{eq:pi21}, one can also see that the Peltier effect is related to the group velocity of the degenerate bands~\cite{suppl}, whereas the Peltier effect by the linearly polarized light is related to the Berry connection of phonons~\cite{Ishizuka2024a}.
The close resemblance of the Peltier effect to the photovoltaics implies that the polarization-dependent Peltier effect is essentially the injection current of phonons.

For further discussion, we focus on the doubly degenerate case as in the Te-like model.
In this case, the equality below Eq.~\eqref{eq:pi21} indicates that $v^\lambda_{n\alpha,n\beta}(\bm 0)=\bar v^{\lambda}_n\;(\sigma^y)_{\alpha\beta}$ where $\bar v^{\lambda}_n$ is a real constant;
note that the same relation holds for the phonon angular momentum, $(L^{\rm ph}_z)_{n\alpha,n\beta}=\bar L^{\rm ph}_{zn}\;(\sigma^y)_{\alpha\beta}$, where $\bar L^{\rm ph}_{zn}$ is a real constant.
The proportionality of operators for phonon energy current and angular momentum indicates $\Pi_{\lambda;\mu\nu}(0;\omega,-\omega)\propto \chi^{(L)}_{\lambda;\mu\nu}(0;\omega,-\omega)$ near the resonance frequency $\omega=\omega_n(\bm0)$. Namely, there is a direct relation between the polarization-dependent Peltier effect and the angular momentum of chiral phonons.

{\it Orbital photocurrent} --- 
The simultaneous induction of angular momentum and the Peltier effect implies that an orbital current of phonons also occurs.
Here, the orbital current refers to the current of phonon angular momentum, analogous to the spin current of electrons.
Intuitively, it is a counter flow of phonons with positive and negative angular momentum.

To study the orbital current, we define the orbital current operator $J_\lambda^{\rm orb}$ from the continuity equation, from which the orbital current and torque terms are defined~\cite{suppl}.
%To investigate the orbital current generation, we calculated the nonlinear conductivity for the orbital current $J_\lambda^{\rm orb}$, whose elements $o_{nm}^{\lambda}(\bm k)$ are
The formula is in the form of Eq.~\eqref{eq:O} with $o^\lambda$ being
\begin{align}
  v_{nm}^{L\lambda}(\bm k)=&\frac{\langle \bm kn|M^z\partial_{k_\nu}\tilde A_{a\lambda,b\sigma}(\bm k)|\bm km\rangle}{4\sqrt{\omega_{\bm kn}\omega_{\bm km}}}\nonumber\\
  &\qquad-\frac{\langle \bm kn|(M^z\tilde I^\nu-\tilde I^\nu M^z)|\bm km\rangle}{8\sqrt{\omega_{\bm kn}\omega_{\bm km}}},\label{eq:Jorb} %\\
  %v_{nm}^{L\lambda}(\bm k)=&\frac{\langle n\bm k|M^z\partial_{k_\lambda}\tilde A(\bm k)|m\bm k\rangle}{4\sqrt{\omega_{n\bm k}\omega_{m\bm k}}},\label{eq:Jorb}
\end{align}
and $v_{nm}^{L\lambda}(\bm k)=v_{n\bar m}^{L\lambda}(\bm k)=v_{\bar nm}^{L\lambda}(\bm k)=v_{\bar n\bar m}^{L\lambda}(\bm k)$, where $\tilde I_{a\lambda,b\sigma}^\nu=\frac{\delta_{ab}}{m_a}\left.\sum_{c}\partial_{k_\nu}A_{a\lambda,c\sigma}(\bm k)\right|_{\bm k=\vec0}$.
%This formula is derived from the continuity equation as discussed in the Supplemental Material~\cite{suppl}.
The numerator of Eq.~\eqref{eq:Jorb} is the product of the angular momentum and the phonon velocity, consistent with the physical intuition.
%whose operator is defined by $J_\lambda^{\rm orb}=\frac12\sum_{\bm k}\psi_{\bm k}^\dagger(\hat v_{\bm k}^\lambda \hat M_{\bm k} + \hat M_{\bm k} \hat v_{\bm k}^\lambda)\psi_{\bm k}$, where $\hat v_{\bm k}^\lambda$ ($\hat M_{\bm k}$) is a $2n_{ph}\times 2n_{ph}$ matrix whose elements are the matrix elements for $J_Q^\lambda$ ($L_z^{\rm ph}$) and $\psi_{\bm k}=(b_{1\bm k},\cdots,b_{n_{ph}\bm k},b_{1\bm k}^\dagger,\cdots,b_{n_{ph}\bm k}^\dagger)^t$ is a vector of the phonon annihilation and creation operators with momentum $\bm k$.
Orbital current generation is studied by combining the nonlinear response theory and Eq.~\eqref{eq:Jorb}.

\begin{figure}
  \includegraphics[width=\linewidth]{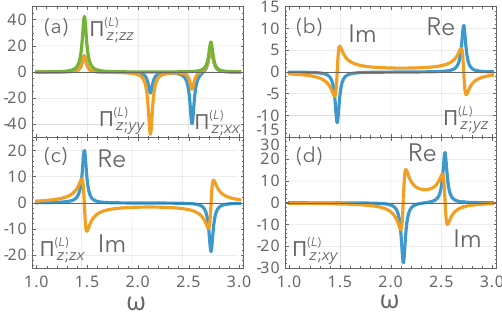}
  \caption{
  (a) Real part of the orbital current conductivity $\Pi^{(L)}_{z;xx}(0;\omega,-\omega)$, $\Pi^{(L)}_{z;yy}(0;\omega,-\omega)$, and $\Pi^{(L)}_{z;zz}(0;\omega,-\omega)$, and the frequency dependence of (b) $\Pi^{(L)}_{z;yz}(0;\omega,-\omega)$, (c) $\Pi^{(L)}_{z;zx}(0;\omega,-\omega)$, and (d) $\Pi^{(L)}_{z;xy}(0;\omega,-\omega)$ for the Te-like model. The results are for $\delta=0.23$, $K_1=3/2$, $K_2=1/2$, $K_3=3$, $\theta=
 0.3\pi$, $a=1$, $c=1$, $m=1$, $q_1=0$, $q_2=1$, $q_3=-1$, and $\tau=20$.
  }\label{fig:piS}
\end{figure}

The calculated conductivities of the nonlinear orbital current $\Pi^{(L)}_{\lambda;\mu\nu}(0;\omega,-\omega)$ are shown in Fig.~\ref{fig:piS}.
Unlike the cases for the angular momentum and Peltier effect, in all elements of the conductivity tensor, the resonance peaks appear in the real part of the conductivity.
The peaks in $\text{Re}[\Pi^{(L)}]$ for chiral phonons ($\omega\sim1.47$ and $2.72$) show that the orbital current occurs with linearly-polarized light while neither angular momentum nor the Peltier effect occurs. %; it is a pure orbital current.
This result is intuitively understandable considering that a linearly polarized light can be viewed as the superposition of right- and left-polarized lights.
The right polarized light induces a phonon current with a non-zero phonon angular momentum and the left polarized light induces an opposite phonon current with the opposite angular momentum; the sum of the two results in the orbital current.
Indeed, the sign of $\Pi^{(L)}_{\lambda;\mu\nu}$ matches that of the product of $\chi^{(L)}_{\lambda;\mu\nu}$ and $\Pi_{\lambda;\mu\nu}$.
The result indicates that chiral phonons can induce a pure orbital current without phonon current and angular momentum.

We note that non-chiral phonons also contributes to the orbital current as in Fig.~\ref{fig:piS}(a).
However, in this case, the linearly polarized light also induces orbital angular momentum and Peltier effect; hence, not a pure orbital current.
Intuitively, it shows that if a phonon with nonzero angular momentum flows, it also induces an orbital current.
%In other words, the orbital current for non-chiral phonons is a polarized phonon current whose angular momentum is nonzero.

{\it Summary} --- 
In this work, we studied the nonlinear responses of chiral phonons to electromagnetic waves: phonon angular-momentum generation, the Peltier effect of phonons, and the photo-induced phonon orbital current.
We find that the angular momentum of the phonon is generated by both chiral and non-chiral phonons.
A unique feature of chiral phonons is a polarization-dependent contribution proportional to $\tau^2$; this behavior contrasts with the contribution from non-chiral phonons, which is proportional to $\tau^1$.
A similar behavior is also obtained for the Peltier effect of phonons, in which we find a new mechanism unique to the chiral phonons.
In addition, we find that the angular momentum and the Peltier effect by chiral phonons are linearly proportional to each other.
Lastly, we argued that an orbital photocurrent of phonons, the angular-momentum current similar to the spin current, occurs by the light irradiation.
The order estimates of these phenomena indicate that chiral phonons realize larger thermal and orbital responses~\cite{suppl}, which facilitates the study of the phenomena and their applications.
%These results uncover rich optical response unique to the chiral phonons and their mutual relation.

\acknowledgements
This work is supported by JSPS KAKENHI (Grants No. JP23K03275 and JP25H00841), JST PRESTO (Grant No. JPMJPR2452), and JST CREST (Grant No. JPMJCR24R5).

%\printbibliography % print bibliography using biblatex
%\bibliography{ref} % print bibliography using natbib


\end{document}

\preprint{APS/123-QED}

\title{
Supplemental Material to \\
``{\it Nonlinear optical response of truly chiral phonons:\\Light-induced phonon angular momentum, Peltier effect, and orbital current}''
}

\author{Hiroaki Ishizuka}
\affiliation{
Department of Physics, Institute of Science Tokyo, Meguro, Tokyo, 152-8551, Japan
}

\author{Masahiro Sato}
\affiliation{
Department of Physics, Chiba University, Chiba 263-8522, Japan
}

\date{\today}

% \begin{abstract}
% \end{abstract}

\pacs{
}% PACS, the Physics and Astronomy

\maketitle
\onecolumngrid
%%%%%%%%%%%%%%%%%%%%%%%%%%%%%%%%%%%%%%%%%%%%%
%%%%%%%%%%%%%%%%%%%%%%%%%%%%%%%%%%%%%%%%%%%%%
\section{Nonlinear response theory}

To study the second-order dc response of quantum phonons to electromagnetic waves, we use the nonlinear response theory.
Phenomenologically, the expectation value of observable $\hat O_\lambda$ induced by the second-order response reads
\begin{align}
    O_\lambda(\Omega)=\sum_{\mu,\nu}\int\frac{d\omega}{2\pi}\chi_{\lambda;\mu\nu}(\Omega;\omega,\Omega-\omega)E_\mu(\omega)E_\nu(\Omega-\omega),\label{eq:current}
\end{align}
where $E_\mu(\omega)$ is the amplitude of the electric field along $\mu$ axis with frequency $\omega$, $\lambda$ is an index such as the direction of magnetization, current, etc., and $\chi_{\lambda;\mu\nu}(\Omega;\omega,\Omega-\omega)$ is the second-order response tensor.

In this work, we consider non-interacting phonon models with $n_{ph}$ phonon bands whose Hamiltonian is given by
\begin{align}
\hat H=\sum_{\substack{\bm k\\n=1,\cdots,n_{ph}}}\omega_{n\bm k}(\hat b_{n\bm k}^\dagger\hat b_{n\bm k}+\frac12),
\end{align}
and the light-matter coupling,
\begin{align}
\hat H'=&\sum_{\mu,n,\bm k}[\beta^\mu_{n\bm k}\hat b_{n\bm k}+(\beta^\mu_{n\bm k})^\ast\hat b_{n\bm k}^\dagger] E_\mu(t),\label{eq:H'}
%\hat H'=&\sum_\mu\hat B^\mu E_\mu(t),\label{eq:H'}\\
%\hat B^\mu=&\sum_{n,\bm k}\beta^\mu_{n\bm k}\hat b_{n\bm k}+(\beta^\mu_{n\bm k})^\ast\hat b_{n\bm k}^\dagger,
\end{align}
with $\beta^\mu_{n\bm k}$ being the coupling constant and $E^\mu(t)$ is the external field.
For this setup, the dc response tensor, $\chi_{\lambda;\mu\nu}(0;\omega,-\omega)$, reads~\cite{Ishizuka2024a}
\begin{align}
\chi_{\lambda;\mu\nu}(0;\omega,-\omega)=&
-\frac1{2\pi V}\sum_{\bm k,n,m}\frac{\beta_{n\bm k}^\mu[o^\lambda_{n\bar m}(\bm k)+o^\lambda_{m\bar n}(-\bm k)]\beta_{m,-\bm k}^\nu}{(\omega-\omega_{n\bm k}-\frac{\rm i}{2\tau})(\omega_{n\bm k}+\omega_{m\bm k}+\frac{\rm i}{2\tau})}+\frac1{2\pi V}\sum_{\bm k,n,m}\frac{(\beta_{n\bm k}^\mu)^\ast[o^\lambda_{\bar nm}(-\bm k)+o^\lambda_{\bar mn}(\bm k)](\beta_{m,-\bm k}^\nu)^\ast}{(\omega+\omega_{n\bm k}-\frac{\rm i}{2\tau})(\omega_{n\bm k}+\omega_{m\bm k}-\frac{\rm i}{2\tau})}\nonumber\\
  &+\frac1{2\pi V}\sum_{\bm k,n,m}\frac{\beta_{n\bm k}^\mu[o^\lambda_{nm}(\bm k)+o^\lambda_{\bar m\bar n}(-\bm k)](\beta_{m\bm k}^\nu)^\ast}{(\omega-\omega_{n\bm k}-\frac{\rm i}{2\tau})(\omega_{n\bm k}-\omega_{m\bm k}+\frac{\rm i}{2\tau})}-\frac1{2\pi V}\sum_{\bm k,n,m}\frac{(\beta_{n\bm k}^\mu)^\ast[o^\lambda_{mn}(\bm k)+o^\lambda_{\bar n\bar m}(-\bm k)]\beta_{m\bm k}^\nu}{(\omega+\omega_{n\bm k}-\frac{\rm i}{2\tau})(\omega_{n\bm k}-\omega_{m\bm k}-\frac{\rm i}{2\tau})}.\label{eq:chi_full}
\end{align}
Here, $\hat b_{n\bm k}$ ($\hat b_{n\bm k}^\dagger$) is the annihilation (creation) operator of a phonon with band index $n$ and momentum $\bm k$, $\omega_{n\bm k}$ ($\omega_{n\bm k}\le \omega_{m\bm k}$ if $n<m$) is the phonon frequency, $\tau=\tau(T)$ is the phenomenological phonon lifetime at temperature $T$, and $o^\lambda_{nm}(\bm k)$ are the matrix elements of the physical quantity,
\begin{align}
\hat O_\lambda=&\sum_{n,m,\bm k}\hat b_{n\bm k}^\dagger o^\lambda_{nm}(\bm k)\hat b_{m\bm k}+\hat b_{n\bm k}^\dagger o^\lambda_{n\bar m}(\bm k)\hat b_{m-\bm k}^\dagger
+\hat b_{n-\bm k} o^\lambda_{\bar nm}(\bm k)\hat b_{m\bm k}+\hat b_{n-\bm k} o^\lambda_{\bar n\bar m}(\bm k)\hat b_{m-\bm k}^\dagger.
%\hat O_\lambda&=\sum_{n,m,\bm k}\hat b_{n\bm k}^\dagger o^\lambda_{nm}(\bm k)\hat b_{m\bm k}+\sum_{n,m,\bm k}\hat b_{n\bm k}^\dagger o^\lambda_{n\bar m}(\bm k)\hat b_{m-\bm k}^\dagger\nonumber\\
%&+\sum_{n,m,\bm k}\hat b_{n-\bm k} o^\lambda_{\bar nm}(\bm k)\hat b_{m\bm k}+\sum_{n,m,\bm k}\hat b_{n-\bm k} o^\lambda_{\bar n\bar m}(\bm k)\hat b_{m-\bm k}^\dagger.\label{eq:O}
\end{align}
In general, the relaxation time scales $\tau\propto\omega$ in the boson cases~\cite{Abrikosov1963a}. However, we take $\tau$ to be a constant of the frequency as it does not affect the argument in this work; an extention to the $\tau\propto\omega$ case is straight forward~\cite{Ishizuka2019a}.

%%%%%%%%%%%%%%%%%%%%%%%%%%%%%%%%%%%%%%%%%%%%%
%%%%%%%%%%%%%%%%%%%%%%%%%%%%%%%%%%%%%%%%%%%%%
\section{Phonon angular momentum}

In classical theory, the total angular momentum of a many-particle system is given by $\vec L=\sum_{l}m_l\vec r_{l}\times\dot{\vec r}_{l}.$
Here, $\vec r_{l}$ is the position of $l$ th particle, $\dot{\vec r}_{l}=d{\vec r}_{l}/dt$, and $m_l$ is the mass of the $l$ th particle.
For the case of phonons~\cite{Vonsovskii1962a}, in which the atoms vibrate around their equilibrium position, the angular momentum reads
\begin{align}
    \vec L=\vec L_{\rm lat}+\vec L_{\rm ph},\quad \vec L_{\rm lat}=\sum_{l\alpha} m_\alpha\vec r_{l\alpha}^0\times\dot{\vec u}_{l\alpha},\quad \vec L_{\rm ph}=\sum_{l\alpha} m_\alpha\vec u_{l\alpha}\times\dot{\vec u}_{l\alpha}. \label{eq:Jph}
\end{align}
Here, we defined the displacement $\vec u_{l\alpha}$ of $\alpha$ the sublattice in the $l$ th unit cell by $\vec r_{l\alpha}=\vec r_{l\alpha}^0+\vec u_{l\alpha}$, where $\vec r_{l\alpha}^0$ is the equilibrium position;
in Eq.~\eqref{eq:Jph}, we used the subscripts $l\alpha$ instead of $l$ to explicitly denote the sublattice.
Among the two terms, $\vec L^{\rm lat}=\sum_{l\alpha} m_\alpha\vec r_{l\alpha}^0\times\dot{\vec u}_{l\alpha}$ corresponds to the rotation of the lattice and $\vec L^{\rm ph}=\sum_{l\alpha} m_\alpha\vec u_{l\alpha}\times\dot{\vec u}_{l\alpha}$ is the phonon (pseudo) angular momentum.

Following the standard protocol for quantizing phonons~\cite{Grosso2013a}, we can obtain
\begin{align}
L^\lambda_{\rm ph}=
&\frac\hbar4\sum_{a,\vec k,n,n'}\frac{\omega_{n\vec k}+\omega_{n'\vec k}}{\sqrt{\omega_{n\vec k}\omega_{n'\vec k}}}\langle n\vec k|M^\lambda|n'\vec k\rangle(b_{n\vec k}^\dagger b_{n'\vec k}-b_{-n\vec k}b_{-n'\vec k}^\dagger)\nonumber\\
%&+i\frac\hbar4\sum_{a,\vec k,n,n'}\frac{-\omega_{n'\vec k}+\omega_{n\vec k}}{\sqrt{\omega_{n\vec k}\omega_{n'\vec k}}}\left[\langle n'\vec k|_{ay}|n\vec k\rangle_{ax}-\langle n'\vec k|_{ax}|n\vec k\rangle_{ay}\right] (b_{-n\vec k}b_{n'\vec k}- b_{n\vec k}^\dagger b_{-n'\vec k}^\dagger).\\
&+\frac\hbar4\sum_{a,\vec k,n,n'}\frac{\omega_{n\vec k}-\omega_{n'\vec k}}{\sqrt{\omega_{n'\vec k}\omega_{n\vec k}}}\langle n\vec k|M^\lambda|n'\vec k\rangle(b_{n\vec k}^\dagger b_{-n'\vec k}^\dagger-b_{-n\vec k}b_{n'\vec k}).
\end{align}
Note that in this work, we considered not only the terms in the first line but also the second line, which was ignored in a previous work~\cite{Zhang2014a}.
The idea behind ignoring the second line was that they are fastly oscillating and, hence, do not contribute to the measurement.
However, as we deal with optical transitions, these terms may also contribute.

Strictly speaking, $\bm L_{\rm ph}$ is not a conserved quantity in a crystal, as the lattice breaks the rotational symmetry.

\section{Flux and torque of the angular momentum}\label{sec:orbital-current}

To study the flow of angular momentum, we focus on the orbital current of atoms, similar to the spin current of electrons.
The argument here is similar to that of the energy current developed by Hardy~\cite{Hardy1963a}.
To define the orbital current from the continuity equation, we consider the angular momentum at position $\bm x$ defined by
\begin{align}
    \bm L_{\rm atm}(\bm x)=\sum_{ia}\Delta(\bm x-\bm r_{i})\bm r_{i}\times\bm p_{i},\label{eq:Latm}
\end{align}
where $\bm r_{i}=(x_i,y_i,z_i)$ [$\bm p_{i}=(p_{ix},p_{iy},p_{iz})$] is the position [momentum] operator of the $i$th atom, and $\Delta(\bm x)$ is an analytic function localized around $\bm x=\bm0$ and $\int\Delta(\bm r) dr^d=1$.

To be concrete, we consider the Hamiltonian
\begin{align}
H=\sum_{i}\frac{\bm p_{i}^2}{2m_i}+V_i(\{\bm r_{l}\}).\label{eq:hamil0}
\end{align}
Here, $p_{i\mu}=-i\hbar\partial_{r_{i\mu}}$, and $V_{i}(\{\bm r_{l}\})$ are the potentials, which we assume to be the function of $\{\bm r_{l}\}$, and satisfies the following conditions,
\begin{enumerate}
    \item $\partial_{r_{j\mu}} V_{i}(\{\bm r_{l}\}) =-\partial_{r_{i\mu}} V_{j}(\{\bm r_{l}\})$ ($i\ne j$),
    \item $\displaystyle\partial_{r_{i\mu}} V_{i}(\{\bm r_{l}\}) =\sum_{j\ne i}\partial_{r_{i\mu}} V_{j}(\{\bm r_{l}\})$.
\end{enumerate}
For example, when the potential $V_{i}(\{\bm r_{l}\})$ is given by a two-particle interaction $v(\bm r_{i},\bm r_{j})$ satisfying the law of action and reaction, $\partial_{r_\mu}v(\bm r,\bm r')=-\partial_{r'_\mu}v(\bm r,\bm r')$, one can define
\begin{align}
    V_{i}=\frac12\sum_{j\ne i}v(\bm r_{i},\bm r_{j}). \label{eq:2p-int}
\end{align}
In this case, $\bm F_{i;j}=-2\nabla_{r_{i}} V_{j}=-\nabla_{r_{i}} v(\bm r_{i},\bm r_{j})$ ($i\ne j$) gives the force acting on $i$th particle by the interaction with $j$th particle.
%Hence, the condition $\partial_{r_{j\mu}} V_{i}(\{\bm u_{l}\}) =-\partial_{r_{i\mu}} V_{j}(\{\bm u_{l}\})$ corresponds to the law of action and reaction.
%Note that, although we use an affinity notation for the phonons, the derivation here and below also applies to non-crystalline systems.

We study the dynamics of $\bm L_{\rm atm}(\bm x)$ using Heisenberg's equation of motion,
\begin{align}
    \partial_tL_{\rm atm}^\mu(\bm x)=-\frac i\hbar[L_{\rm atm}^\mu(\bm x),H].
    \label{eq:Heisen0}
\end{align}
By substituting Eqs.~\eqref{eq:Latm} and \eqref{eq:hamil0} into Eq.~\eqref{eq:Heisen0}, we find
\begin{align}
    \partial_tL_{\rm atm}^\mu(\bm x)=&-\frac i\hbar\sum_{\substack{i,j\\\nu\lambda}}\left[\Delta(\bm x-\bm r_{i})\epsilon_{\mu\nu\lambda}r_{i\nu}p_{i\lambda},\frac{\bm p_{j}^2}{2m_b}\right]-\sum_{\substack{i,j\\\nu\lambda}}\Delta(\bm x-\bm r_{i})\epsilon_{\mu\nu\lambda}r_{i\nu}\partial_{r_{i\lambda}}V_{j},\\
    =&-\sum_{\substack{i,j\\\nu\lambda\rho}}\frac i{m_i}\partial_{\rho}\Delta(\bm x-\bm r_{i})\epsilon_{\mu\nu\lambda}r_{i\nu}p_{i\lambda}p_{i\rho}+\sum_{\substack{i,j\\\nu\lambda\rho}}\frac{i\hbar}{2m_i}\partial_{\rho}^2\Delta(\bm x-\bm r_{i})\epsilon_{\mu\nu\lambda}r_{i\nu}p_{i\lambda}\nonumber\\
    &+\sum_{\substack{i,j\\\nu\lambda}}\frac{i\hbar}{m_i}\epsilon_{\mu\nu\lambda}\partial_{\nu}\Delta(\bm x-\bm r_{i})p_{i\lambda}-\sum_{\substack{i\ne j\\\nu\lambda}}\left[\Delta(\bm x-\bm r_{i})-\Delta(\bm x-\bm r_{j})\right]\epsilon_{\mu\nu\lambda}r_{i\nu}\partial_{r_{i\lambda}}V_{j}+T^\mu(\bm x),\label{eq:heis0}
    %&\hspace{20mm}-\frac{\Delta(\bm x-\bm r_{i})+\Delta(\bm x-\bm r_{j})}2\epsilon_{\mu\nu\lambda}r_{i\nu}\partial_{r_{i\lambda}}V_{j}\label{eq:heis}\\
\end{align}
where
\begin{align}
    T^\mu(\bm x)=&-\sum_{\substack{i\ne j\\\nu\lambda}}\left[\Delta(\bm x-\bm r_{i})+\Delta(\bm x-\bm r_{j})\right]\epsilon_{\mu\nu\lambda}r_{i\nu}\partial_{r_{i\lambda}}V_{j}.
    \label{eq:heis0_1}
\end{align}
In Eq.~\eqref{eq:heis0}, the first three terms on the right-hand side are the change of angular momentum by the motion of particles.
On the other hand, the last two terms are the transfer and generation of angular momentum by the interaction.

To understand the meaning of the fourth term, let us consider a rotationally symmetric system with a two-particle interaction, in which the angular momentum is conserved.
In this case, the force acting between the $i$th and $j$th particles should be $\bm F_{i;j}=\frac{\bm r_i-\bm r_j}{|\bm r_i-\bm r_j|}\tilde F_{i;j}$ and $\tilde F_{i;j}=\tilde F_{j;i}$.
By substituting this equation into Eqs.~\eqref{eq:heis0} and \eqref{eq:heis0_1}, one can show that $T^\mu(\bm x)=0$ and Eq.~\eqref{eq:heis0} reads
\begin{align}
    \partial_tL_{\rm atm}^\mu(\bm x)
    =&-\sum_{\substack{i,j\\\nu\lambda\rho}}\frac i{m_i}\partial_{\rho}\Delta(\bm x-\bm r_{i})\epsilon_{\mu\nu\lambda}r_{i\nu}p_{i\lambda}p_{i\rho}+\sum_{\substack{i,j\\\nu\lambda\rho}}\frac{i\hbar}{2m_i}\partial_{\rho}^2\Delta(\bm x-\bm r_{i})\epsilon_{\mu\nu\lambda}r_{i\nu}p_{i\lambda}\nonumber\\
    &+\sum_{\substack{i,j\\\nu\lambda}}\frac{i\hbar}{m_i}\epsilon_{\mu\nu\lambda}\partial_{\nu}\Delta(\bm x-\bm r_{i})p_{i\lambda}+\sum_{\substack{i\ne j\\\nu\lambda}}\Delta(\bm x-\bm r_{i})(\bm r_{i}\times\bm r_j)_\mu \tilde F_{i;j}.
\end{align}
The last term in this equation corresponds to the fourth term in Eq.~\eqref{eq:heis0}.
The fact that the fourth term survives in the rotationally symmetric case indicates that it corresponds to the transfer of angular momentum through the interaction.

The fact that $T^\mu(\bm x)$ vanishes in the rotationally symmetric case suggests that it is a torque term that violates the angular momentum conservation.
To see this more explicitly, let us consider the time evolution of total angular momentum.
By taking the spatial integral of Eq.~\eqref{eq:heis0}, the first three terms are the surface terms which we assume to be negligible, and the fourth term vanishes; the vanishing fourth term indicates that it does not break the conservation of the angular momentum.
Therefore, only the fifth term in the right-hand side remains and the equation of motion for total angular momentum becomes
\begin{align}
    \partial_tL_{\rm atm}^\mu=\int dx^d\partial_tL_{\rm atm}^\mu(\bm x)=\int dx^dT^\mu(\bm x)
    =\sum_{i\ne j}\left(\bm r_{i}\times\bm F_{i;j}\right)_\mu,
\end{align}
where the summation is over all pairs of $(i,j)$ and $\nabla_{r_{i}}=(\partial_{x_{i}},\partial_{y_{i}},\partial_{z_{i}})$.
As $\bm F_{i;j}$ is the effective interaction between the $i$th and $j$th particles, $\bm u_{i}\times\bm F_{i;j}$ is the temporal change of the phonon angular momentum of the $i$th particle.
Therefore, $T^\mu$ is a torque density that violates the angular momentum conservation.

By Taylor expanding $\Delta(\bm x-\bm r_{i})$, 
\begin{align}
    \Delta(\bm x-\bm r_{jb})=\prod_{\nu}\left(\sum_{n=0}^\infty\frac1{n!}(r_{i\nu}-r_{jb\nu})^n\partial_{\nu}^n\right)\Delta(\bm x-\bm r_{i}),
\end{align}
$\partial_tL_{\rm ph}^\mu(\bm x)$ reads
%\begin{align}
%    &\partial_tL_{\rm ph}^\mu(\bm x)+\sum_{\substack{i,jb\\\nu\lambda\rho}}\frac i{m_a}\partial_{x_\rho}\Delta(\bm x-\bm r_{i})\epsilon_{\mu\nu\lambda}u_{i\nu}p_{i\lambda}p_{i\rho}+\frac{i\hbar}{2m_a}\partial_{x_\rho}^2\Delta(\bm x-\bm r_{i})\epsilon_{\mu\nu\lambda}u_{i\nu}p_{i\lambda}-\sum_{\substack{i,jb\\\nu\lambda}}\frac{i\hbar}{m_a}\epsilon_{\mu\nu\lambda}\partial_{x_\nu}\Delta(\bm x-\bm r_{i})p_{i\lambda}\nonumber\\
%    &-\frac12\sum_\rho\sum_{\substack{i,jb\\\nu\lambda}}\left(\sum_{n=1}^\infty\frac1{n!}(r_{i\rho}-r_{jb\rho})^n\partial_{\rho}^n\right)\left[1+\frac12\sum_{\substack{n=1\\\sigma\ne\rho}}^\infty\frac1{n!}(r_{i\sigma}-r_{jb\sigma})^n\partial_{\sigma}^n+\frac13\prod_{\sigma\ne\rho}\left(\sum_{n=1}^\infty\frac1{n!}(r_{i\rho}-r_{jb\rho})^n\partial_{\rho}^n\right)\right]\epsilon_{\mu\nu\lambda}u_{i\nu}\partial_{u_{i\lambda}}V_{jb}\nonumber\\
%    &=-\frac12\left[1+\prod_{\nu}\left(\sum_{n=0}^\infty\frac1{n!}(r_{i\nu}-r_{jb\nu})^n\partial_{\nu}^n\right)\right]\Delta(\bm x-\bm r_{i})\epsilon_{\mu\nu\lambda}u_{i\nu}\partial_{u_{i\lambda}}V_{j}.
%\end{align}
\begin{align}
    &\partial_tL_{\rm ph}^\mu(\bm x)+\sum_\nu \partial_\nu J_{\rm ph}^{\mu\nu}(\bm x)=T^\mu(\bm x),\label{eq:continuity0}
\end{align}
where
\begin{align}
    &J_{\rm ph}^{\mu\nu}(\bm x)=\sum_{\substack{i,j\\\rho\lambda}}\frac i{m_i}\Delta(\bm x-\bm r_{i})\epsilon_{\mu\rho\lambda}u_{i\rho}p_{i\lambda}p_{i\nu}
    +\frac{i\hbar}{2m_i}\partial_{x_\nu}\Delta(\bm x-\bm r_{i})\epsilon_{\mu\rho\lambda}r_{i\rho}p_{i\lambda}
    -\sum_{\substack{i,j\\\lambda}}\frac{i\hbar}{m_i}\epsilon_{\mu\nu\lambda}\partial_\nu\Delta(\bm x-\bm r_{i})p_{i\lambda}\nonumber\\
    &-\sum_{\substack{i\ne j\\\rho\lambda}}\left(\sum_{n=1}^\infty\frac1{n!}(r_{i\nu}-r_{j\nu})^n\partial_{\nu}^{n-1}\right)\left[1+\frac12\sum_{\substack{n=1\\\sigma\ne\nu}}^\infty\frac1{n!}(r_{i\sigma}-r_{j\sigma})^n\partial_{\sigma}^n+\frac13\prod_{\sigma\ne\nu}\left(\sum_{n=1}^\infty\frac1{n!}(r_{i\sigma}-r_{j\sigma})^n\partial_{\sigma}^n\right)\right]\Delta(\bm x-\bm r_{i})\nonumber\\
    &\hspace{20mm}\times\epsilon_{\mu\rho\lambda}r_{i\rho}\partial_{r_{i\lambda}}V_{j},\label{eq:Jorbx}
\end{align}
and
\begin{align}
    T^\mu(\bm x)=-\left[1+\prod_{\nu}\left(\sum_{n=0}^\infty\frac1{n!}(r_{i\nu}-r_{j\nu})^n\partial_{\nu}^n\right)\right]\Delta(\bm x-\bm r_{i})\epsilon_{\mu\nu\lambda}r_{i\nu}\partial_{r_{i\lambda}}V_{j}.
\end{align}
Equation~\eqref{eq:continuity0} should be compared to the standard form of the continuity equation, $\partial_t\rho(\bm x)+\nabla\cdot\bm J(\bm x)=0$, where $\rho(\bm x)$ and $\bm J(\bm x)$ are respectively the charge and flux densities.
As the continuity equation should hold when the angular momentum is conserved, i.e. $T^\mu=0$, we define $J_{\rm ph}^{\mu\nu}$ as the orbital current or the flux of the angular momentum.

Taking the spatial average over Eq.~\eqref{eq:Jorbx}, the average orbital current density reads
\begin{align}
    &J_{\rm ph}^{\mu\nu}=\frac1V\sum_{i,j,\rho,\lambda}\frac i{m_i}\epsilon_{\mu\rho\lambda}r_{i\rho}p_{i\lambda}p_{i\nu}-\frac1{V}\sum_{\substack{i\ne j\\\rho,\lambda}}(r_{i\nu}-r_{j\nu})\epsilon_{\mu\rho\lambda}r_{i\rho}\partial_{r_{i\lambda}}V_{j}.
\end{align}
Here, $V$ is the volume of the system, and we neglected the surface terms.
The first term is the contribution of a moving particle with angular momentum $\bm r_i\times\bm p_i$ and the second term is the transfer of angular momentum through the interaction.

%%%%%%%%%%%%%%%%%%%%%%%%%%%%%%%%%%%%%
%%%%%%%%%%%%%%%%%%%%%%%%%%%%%%%%%%%%%
%%%%%%%%%%%%%%%%%%%%%%%%%%%%%%%%%%%%%
\section{Flux of phonon angular momentum}

%%%%%%%%%%%%%%%%%%%%%%%%%%%%%%%%%%%%%
%%%%%%%%%%%%%%%%%%%%%%%%%%%%%%%%%%%%%
%%%%%%%%%%%%%%%%%%%%%%%%%%%%%%%%%%%%%
\subsection{Orbital current of phonon angular momentum}

Next, we apply a similar method in Sec.~\ref{sec:orbital-current} to the phonon angular momentum and derive the phonon orbital current.
We consider the local phonon angular momentum at $\bm x$ defined by
\begin{align}
    \bm L_{\rm ph}(\bm x)=\sum_{ia}\Delta(\bm x-\bm r_{ia})\bm u_{ia}\times\bm p_{ia},\label{eq:Lph}
\end{align}
where $\bm u_{ia}=\bm r_{ia}-\bm r_{ia}^0$ is the displacement of $a$th sublattice atom in the $i$th unit cell and $\bm r_{ia}^0$ is its equilibrium position.
Note that the phonon angular momentum $\bm L_{\rm ph}(\bm x)$ can be viewed as a generalization of the angular momentum as it is equivalent to the angular momentum of the system if $\bm r_{ia}^0=\bm 0$.
Hence, we can construct the argument similar to that of $\bm L_{\rm atm}(\bm x)$.

\if0
To be concrete, we consider the Hamiltonian
\begin{align}
H=\sum_{ia}\frac{\bm p_{ia}^2}{2m_a}+V_{ia}(\{\bm r_{lc}\}),\label{eq:hamil}
\end{align}
where $\bm p_{ia}=(p_{iax},p_{iay},p_{iaz})$  ($p_{ia
\mu}=-i\hbar\partial_{r_{ia\mu}}$) is the momentum operator for the $a$th sublattice atom in $i$th unit cell, and $V_{ia}(\{\bm r_{lc}\})$ is the $ia$ component of the potential, which we assume to be the function of $\bm r_{lc}$, and $\partial_{r_{j\mu}} V_{ia}(\{\bm r_{lc}\}) =-\partial_{r_{ia\mu}} V_{mb}(\{\bm r_{la}\})$.
For example, in the case of particles with a two-particle interaction $v(\bm r_{ia},\bm r_{j})$, one can define
\begin{align}
    V_{ia}=\frac12\sum_{jb\ne ia}v(\bm r_{ia},\bm r_{jb}). \label{eq:2p-int}
\end{align}
The condition $\partial_{r_{jb\mu}} V_{ia}(\{\bm u_{lc}\}) =-\partial_{r_{ia\mu}} V_{mb}(\{\bm u_{la}\})$ corresponds to the law of action and reaction in the case of two particle interactions.
%Note that, although we use an affinity notation for the phonons, the derivation here and below also applies to non-crystalline systems.
\fi

Using Heisenberg's equation of motion,
\begin{align}
    \partial_tL_{\rm ph}^\mu(\bm x)=-\frac i\hbar[L_{\rm ph}^\mu(\bm x),H],
\end{align}
the time evolution of $\bm L_{\rm ph}(\bm x)$ reads
\begin{align}
    \partial_tL_{\rm ph}^\mu(\bm x)=&-\frac i\hbar\sum_{\substack{ia,jb\\\nu\lambda}}\left[\Delta(\bm x-\bm r_{ia})\epsilon_{\mu\nu\lambda}u_{ia\nu}p_{ia\lambda},\frac{\bm p_{jb}^2}{2m_b}\right]-\sum_{\substack{ia,jb\\\nu\lambda}}\Delta(\bm x-\bm r_{ia})\epsilon_{\mu\nu\lambda}u_{ia\nu}\partial_{r_{ia\lambda}}V_{jb},\\
    =&-\sum_{\substack{ia,jb\\\nu\lambda\rho}}\frac i{m_a}\partial_{\rho}\Delta(\bm x-\bm r_{ia})\epsilon_{\mu\nu\lambda}u_{ia\nu}p_{ia\lambda}p_{ia\rho}+\sum_{\substack{ia,jb\\\nu\lambda\rho}}\frac{i\hbar}{2m_a}\partial_{\rho}^2\Delta(\bm x-\bm r_{ia})\epsilon_{\mu\nu\lambda}u_{ia\nu}p_{ia\lambda}\nonumber\\
    &+\sum_{\substack{ia,jb\\\nu\lambda}}\frac{i\hbar}{m_a}\epsilon_{\mu\nu\lambda}\partial_{\nu}\Delta(\bm x-\bm r_{ia})p_{ia\lambda}-\sum_{\substack{ia\ne jb\\\nu\lambda}}\left[\Delta(\bm x-\bm r_{ia})-\Delta(\bm x-\bm r_{jb})\right]\epsilon_{\mu\nu\lambda}u_{ia\nu}\partial_{r_{ia\lambda}}V_{jb}+T^\mu_{\rm ph}(\bm x),\label{eq:heis}
    %&\hspace{20mm}-\frac{\Delta(\bm x-\bm r_{ia})+\Delta(\bm x-\bm r_{jb})}2\epsilon_{\mu\nu\lambda}u_{ia\nu}\partial_{u_{ia\lambda}}V_{jb}\label{eq:heis}\\
\end{align}
where
\begin{align}
    T^\mu_{\rm ph}(\bm x)=&-\sum_{\substack{ia\ne jb\\\nu\lambda}}\left[\Delta(\bm x-\bm r_{ia})+\Delta(\bm x-\bm r_{jb})\right]\epsilon_{\mu\nu\lambda}u_{ia\nu}\partial_{u_{ia\lambda}}V_{jb}.
\end{align}
In Eq.~\eqref{eq:heis}, the first three terms on the right-hand side are the change of angular momentum by the motion of particles, and the last two terms are the transfer of the angular momentum between particles through the interaction.

To understand the physical meaning of the last two terms in Eq.~\eqref{eq:heis}, let us consider the time evolution of the total phonon angular momentum.
By taking the spatial integral of Eq.~\eqref{eq:heis}, the first three terms are the surface terms, which we assume to be negligible, and the fourth term vanishes; the vanishing fourth term indicates that it does not break the conservation of the angular momentum.
Therefore, only the fifth term in the right-hand side remains and the equation of motion for total angular momentum becomes
\begin{align}
    \partial_tL_{\rm ph}^\mu=\int dx^d T^\mu_{\rm ph}(\bm x)=\sum_{ia\ne jb}\left(\bm u_{ia}\times\bm F_{ia;jb}\right)_\mu,\quad \bm F_{ia;jb}=-2\nabla_{r_{ia}} V_{jb}.
\end{align}
%For the case of two-particle interaction in Eq.~\eqref{eq:2p-int}, $\bm F_{ia;jb}=-\nabla_{r_{ia}} v(\bm r_{ia},\bm r_{jb})$ is the force acting on the particle $ia$ from $jb$.
%Hence, $\bm F_{ia;jb}$ can be considered as an effective interaction between the particles $ia$ and $jb$.
Considering that $\bm u_{ia}\times\bm F_{ia;jb}$ is the temporal change of the phonon angular momentum of the ion $ia$, we find that each term on the right-hand side becomes non-zero only if the induced angular momentum of ion $ia$ is different from the reduction of angular momentum $jb$.
Therefore, $T^\mu$ is a torque density that violates the conservation of the phonon angular momentum.

To see the dynamics of phonon angular momentum in the presence of the conservation law, we consider a one-dimensional chain with continuous rotational symmetry about the chain axis.
For concreteness, let us consider a chain along the $z$ axis, in which $L_{\rm ph}^z$ is a conserved quantity; this assumption does not reduce the generality.
Assuming a small deviation $u_{ia\mu}$ and expanding the potential up to the second order in $u_{ia\mu}$, the potential term reads
\begin{align}
    V_{\rm ph}\simeq \frac12\sum_{(mc,nd)}D^{\mu\nu}_{mc,nd}(u_{mc\mu}-u_{nd\mu})(u_{mc\nu}-u_{nd\nu})+\text{const.},
    \label{eq:Vquad}
\end{align}
where $D^{\mu\nu}_{mc,nd}=D^{\nu\mu}_{mc,nd}=D^{\mu\nu}_{nd,mc}$ is the coupling constant and the sum is over all pairs of atoms; the corresponding $V_{jb}$ is $V_{jb}= \frac14\sum_{mc,\mu,\nu}D^{\mu\nu}_{jb,mc}(u_{mc\mu}-u_{jb\mu})(u_{mc\nu}-u_{jb\nu})+\text{const.}$
The potential is the function of the difference $u_{mc\mu}-u_{jb\mu}$ due to the translational symmetry of atoms, i.e., the potential remains the same when shifting all atoms by the same amount.
The with this potential, the torque reads
\begin{align}
    T^z_{\rm ph}(\bm x)=&-\sum_{ia\ne jb}\left[\Delta(\bm x-\bm r_{ia})+\Delta(\bm x-\bm r_{jb})\right](u_{iax}\partial_{r_{iay}}V_{jb}-u_{iay}\partial_{r_{iax}}V_{jb}),\\
    =&-\frac12\sum_{ia\ne jb}\left[\Delta(\bm x-\bm r_{ia})+\Delta(\bm x-\bm r_{jb})\right]\left[u_{iax}D^{y\nu}_{ia,jb}(u_{ia\nu}-u_{jb\nu})-u_{iay}D^{x\nu}_{ia,jb}(u_{ia\nu}-u_{jb\nu})\right].
\end{align}
With the rotational symmetry, $D_{ia,jb}^{xx}=D_{ia,jb}^{yy}=D_{ia,jb}$ and $D^{xz}_{ia,jb}=D^{yz}_{ia,jb}=D^{zx}_{ia,jb}=D^{zy}_{ia,jb}=0$.
Hence,
\begin{align}
    T^z_{\rm ph}(\bm x)=&\frac12\sum_{ia\ne jb}\left[\Delta(\bm x-\bm r_{ia})+\Delta(\bm x-\bm r_{jb})\right]D_{ia,jb}(\vec u_{ia}\times\vec u_{jb})_z=0.
\end{align}
The torque vanishes in the presence of the rotational symmetry.

By Taylor expanding $\Delta(\bm x-\bm r_{ia})$, 
\begin{align}
    \Delta(\bm x-\bm r_{jb})=\prod_{\nu}\left(\sum_{n=0}^\infty\frac1{n!}(r_{ia\nu}-r_{jb\nu})^n\partial_{\nu}^n\right)\Delta(\bm x-\bm r_{ia}),
\end{align}
$\partial_tL_{\rm ph}^\mu(\bm x)$ reads
%\begin{align}
%    &\partial_tL_{\rm ph}^\mu(\bm x)+\sum_{\substack{ia,jb\\\nu\lambda\rho}}\frac i{m_a}\partial_{x_\rho}\Delta(\bm x-\bm r_{ia})\epsilon_{\mu\nu\lambda}u_{ia\nu}p_{ia\lambda}p_{ia\rho}+\frac{i\hbar}{2m_a}\partial_{x_\rho}^2\Delta(\bm x-\bm r_{ia})\epsilon_{\mu\nu\lambda}u_{ia\nu}p_{ia\lambda}-\sum_{\substack{ia,jb\\\nu\lambda}}\frac{i\hbar}{m_a}\epsilon_{\mu\nu\lambda}\partial_{x_\nu}\Delta(\bm x-\bm r_{ia})p_{ia\lambda}\nonumber\\
%    &-\frac12\sum_\rho\sum_{\substack{ia,jb\\\nu\lambda}}\left(\sum_{n=1}^\infty\frac1{n!}(r_{ia\rho}-r_{jb\rho})^n\partial_{\rho}^n\right)\left[1+\frac12\sum_{\substack{n=1\\\sigma\ne\rho}}^\infty\frac1{n!}(r_{ia\sigma}-r_{jb\sigma})^n\partial_{\sigma}^n+\frac13\prod_{\sigma\ne\rho}\left(\sum_{n=1}^\infty\frac1{n!}(r_{ia\rho}-r_{jb\rho})^n\partial_{\rho}^n\right)\right]\epsilon_{\mu\nu\lambda}u_{ia\nu}\partial_{u_{ia\lambda}}V_{jb}\nonumber\\
%    &=-\frac12\left[1+\prod_{\nu}\left(\sum_{n=0}^\infty\frac1{n!}(r_{ia\nu}-r_{jb\nu})^n\partial_{\nu}^n\right)\right]\Delta(\bm x-\bm r_{ia})\epsilon_{\mu\nu\lambda}u_{ia\nu}\partial_{u_{ia\lambda}}V_{jb}.
%\end{align}
\begin{align}
    &\partial_tL_{\rm ph}^\mu(\bm x)+\sum_\nu \partial_\nu J_{\rm ph}^{\mu\nu}(\bm x)=T^\mu(\bm x),\label{eq:continuity}
\end{align}
where
\begin{align}
    &J_{\rm ph}^{\mu\nu}(\bm x)=\sum_{\substack{ia,jb\\\rho\lambda}}\frac i{m_a}\Delta(\bm x-\bm r_{ia})\epsilon_{\mu\rho\lambda}u_{ia\rho}p_{ia\lambda}p_{ia\nu}
    +\frac{i\hbar}{2m_a}\partial_{x_\nu}\Delta(\bm x-\bm r_{ia})\epsilon_{\mu\rho\lambda}u_{ia\rho}p_{ia\lambda}
    -\sum_{\substack{ia,jb\\\lambda}}\frac{i\hbar}{m_a}\epsilon_{\mu\nu\lambda}\partial_\nu\Delta(\bm x-\bm r_{ia})p_{ia\lambda}\nonumber\\
    &-\sum_{\substack{ia,jb\\\rho\lambda}}\left(\sum_{n=1}^\infty\frac1{n!}(r_{ia\nu}-r_{jb\nu})^n\partial_{\nu}^{n-1}\right)\left[1+\frac12\sum_{\substack{n=1\\\sigma\ne\nu}}^\infty\frac1{n!}(r_{ia\sigma}-r_{jb\sigma})^n\partial_{\sigma}^n+\frac13\prod_{\sigma\ne\nu}\left(\sum_{n=1}^\infty\frac1{n!}(r_{ia\sigma}-r_{jb\sigma})^n\partial_{\sigma}^n\right)\right]\Delta(\bm x-\bm r_{ia})\nonumber\\
    &\hspace{20mm}\times\epsilon_{\mu\rho\lambda}u_{ia\rho}\partial_{u_{ia\lambda}}V_{jb},
\end{align}
and
\begin{align}
    T^\mu(\bm x)=-\left[1+\prod_{\nu}\left(\sum_{n=0}^\infty\frac1{n!}(r_{ia\nu}-r_{jb\nu})^n\partial_{\nu}^n\right)\right]\Delta(\bm x-\bm r_{ia})\epsilon_{\mu\nu\lambda}u_{ia\nu}\partial_{u_{ia\lambda}}V_{jb}.
\end{align}
Comparing Eq.~\eqref{eq:continuity} and the standard form of the continuity equation, $\partial_t\rho(\bm x)+\nabla\cdot\bm J(\bm x)=0$, and considering the fact that $\partial_\nu J_{\rm ph}^{\mu\nu}$ conserves the total angular momentum, we define $J_{\rm ph}^{\mu\nu}$ as the flux of phonon angular momentum.

The average orbital current density reads
\begin{align}
    &J_{\rm ph}^{\mu\nu}=-\frac1{V}\sum_{\substack{ia\ne jb\\\rho\lambda}}(r_{ia\nu}-r_{jb\nu})\epsilon_{\mu\rho\lambda}u_{ia\rho}\partial_{u_{ia\lambda}}V_{jb}+\frac1V\sum_{\substack{ia,jb\\\rho\lambda}}\frac i{m_a}\Delta(\bm x-\bm r_{ia})\epsilon_{\mu\rho\lambda}u_{ia\rho}p_{ia\lambda}p_{ia\nu}.
\end{align}
For the potential in Eq.~\eqref{eq:Vquad}, %to the quadratic order in $u_{ia\mu}$ and $p_{ia\mu}$, 
the orbital current density reads
\begin{align}
    J_{\rm ph}^{\mu\nu}=&-\frac1{2V}\sum_{\substack{ia,jb\\\rho\lambda}}(r_{ia\nu}^0-r^0_{jb\nu})\epsilon_{\mu\rho\lambda}u_{ia\rho}D^{\lambda\sigma}_{ia,jb}(u_{ia\sigma}-u_{jb\sigma}),\\
    =&-\frac1{4V}\sum_{\substack{ia,jb\\\rho\lambda}}(r_{ia\nu}^0-r^0_{jb\nu})\epsilon_{\mu\rho\lambda}(u_{ia\rho}+u_{jb\rho})D^{\lambda\sigma}_{ia,jb}(u_{ia\sigma}-u_{jb\sigma}),
\end{align}
within the quadratic order in $u_{ia\mu}$ and $p_{ia\mu}$. 
This is the orbital current formula we used in this work.

%%%%%%%%%%%%%%%%%%%%%%%%%%%%%%%%%%%%%
%%%%%%%%%%%%%%%%%%%%%%%%%%%%%%%%%%%%%
%%%%%%%%%%%%%%%%%%%%%%%%%%%%%%%%%%%%%
\subsection{Orbital current in phonon representation}

For a crystalline system, the orbital current operator reads
\begin{align}
    J_{\rm ph}^{\mu\nu}
    =&-\frac1{2V}\sum_{\substack{\vec k,a,b\\\rho\lambda\sigma}}u_{-\vec ka\rho}i\left[\epsilon_{\mu\rho\lambda}\partial_{q_\nu}A_{a\lambda,b\sigma}(\vec k)-\frac{\epsilon_{\mu\rho\lambda}I_{a\lambda,b\sigma}-I_{a\rho,b\lambda}\epsilon_{\mu\lambda\sigma}}2\right]u_{\vec kb\sigma},
\end{align}
where $u_{\vec qa\mu}=\frac1{\sqrt N}\sum_i u_{\vec qa\mu}e^{i\vec q\cdot\vec r_{ia}}$, $\bm q=(q_x,q_y,q_z)$ is the wavenumber, and $N$ being the number of unit cells. Here, we used a notation for the interaction term in Eq.~\eqref{eq:Vquad}
\begin{align}
V_{\rm ph}=\frac12\sum_{ia\mu,jb\nu}A_{ia\mu,jb\nu}u_{ia\mu}u_{jb\nu},
\end{align}
and $A_{a\mu,b\nu}(\vec q)=\sum_{j}A_{0a\mu,jb\nu}e^{i\vec q\cdot(\vec r_{jb}-\vec r_{0a})}$ for the sake of consistency with related works~\cite{Hardy1963a,Ishizuka2024a}, and
\begin{align}
    I_{a\lambda,b\sigma}^\nu=\delta_{ab}\left.\sum_{c}\partial_{k_\nu}A_{a\lambda,c\sigma}(\vec k)\right|_{\vec k=\vec0}.
\end{align}

Using phonon eigenstates, $J_{\rm ph}^{\mu\nu}$ reads
\begin{align}
    J_{\rm ph}^{\mu\nu}
    =&-\frac\hbar{2V}\sum_{\substack{\vec k,a,b\\\rho\lambda\sigma}}(b_{n\vec k}^\dagger+b_{n,-\vec k})\langle n\vec k|_{a\rho}i\left[\epsilon_{\mu\rho\lambda}\frac{\partial_{q_\nu}A_{a\lambda,b\sigma}(\vec q)}{2\sqrt{m_am_b\omega_{n-\vec k}\omega_{m\vec k}}}-\frac{\epsilon_{\mu\rho\lambda}I^\nu_{a\lambda,b\sigma}-I^\nu_{a\rho,b\lambda}\epsilon_{\mu\lambda\sigma}}{4\sqrt{m_am_b\omega_{n-\vec k}\omega_{m\vec k}}}\right]|m\vec k\rangle_{b\sigma}(b_{m\vec k}^\dagger+b_{m-\vec k}).
\end{align}
This equation corresponds to the orbital current used in the main text.
By introducing matrices $M^\mu$, whose elements are $M_{a\rho,b\lambda}^\mu=-i\delta_{ab}\epsilon_{\mu\rho\lambda}$, we obtain the expression of the orbital current in the main text.

%%%%%%%%%%%%%%%%%%%%%%%%%%%%%%%%%%%%%
%%%%%%%%%%%%%%%%%%%%%%%%%%%%%%%%%%%%%
%%%%%%%%%%%%%%%%%%%%%%%%%%%%%%%%%%%%%
\section{Order estimate}

Here, we roughly estimate the values of nonlinear response functions for the photo-induced angular momentum, photo-induced Peltier effect, and photo-induced orbital current. 
To this end, we assume the following parameter set: the charges are $q_1=0$ and $q_2=-q_3\sim 1.602\times10^{-19}$ C, the atom mass is  $m=127.6$ Da, and the lattice parameters are $a=b=4.45$ $\AA$ 
and $c=5.93$ $\AA$.
In this setup, the hight of the peaks in Figs.~\ref{main:fig:chi}, \ref{main:fig:pi}, and \ref{main:fig:piS} are estimated as $\chi^{(2)}\sim 10^{-15}$ $\hbar$m$^{-2}$V$^{-2}$ per a unit cell, $\Pi\sim10^{-5}$-$10^{-6}$ W/V$^2$, and $\Pi^{(L)}\sim 10^{-17}$-$10^{-18}$ JV$^{-2}$ for chiral phonons. 

The value of $\Pi$ is similar or somewhat larger than a previous work studying non-degenerate phonon modes~\cite{Ishizuka2024a}.
This is naturally expected as chiral phonons tend to give a larger response as in Fig.~\ref{main:fig:pi}.
The magnitude of $\Pi^{(L)}\sim 10^{-17}$-$10^{-18}$ JV$^{-2}$ is also expected to be a few orders of magnitude larger than the spin current by magnetic excitations for the same incident light~\cite{Ishizuka2019a,Ishizuka2019b,Ishizuka2022a}.
Therefore, one can expect larger optical dc responses of chiral phonons in clean enough materials.

\newcommand{\bibitemS}[1]{
\stepcounter{count}
\bibitem[S\thecount]{#1}}
\newcounter{count}
%%%%%%%%%%%%%%%%%%
%
%\bibliography{ref} % print bibliography using natbib
merlin.mbs apsrev4-1.bst 2010-07-25 4.21a (PWD, AO, DPC) hacked